\documentclass[aps,prl,preprintnumbers,amsmath,amssymb,latexsym,nofootinbib,array,enumerate,letter,twocolumn,superscriptaddress]{revtex4-1}
\usepackage{blindtext}
\usepackage{lineno}
\usepackage{bm,color,xcolor}
\usepackage{slashed} 
\usepackage{graphicx}
\usepackage{subfigure} 
\usepackage{amsmath}
\usepackage{amssymb}
\usepackage{multirow}
\usepackage{float}
\usepackage{comment}
\usepackage{fontawesome} \usepackage{mathtools}
\usepackage[colorlinks=true
,urlcolor=blue
,anchorcolor=blue
,citecolor=blue 
,filecolor=blue
,linkcolor=blue
,menucolor=blue
,linktocpage=true
,pdfproducer=medialab
,pdfa=true
]{hyperref}
\usepackage{lipsum}

\makeatletter
\def\l@subsubsection#1#2{}
\makeatother

\newcommand{\MeV}{\ensuremath{\mathrm{MeV}}}

\newcommand{\brr}[1]{\left(#1\right)}
\newcommand{\srr}[1]{\left[#1\right]}

\begin{document}


\title{QCD Axion Domain Walls from Super-Cooling First Order Phase Transition}

\author{Kun-Feng Lyu} 
\email{kunfeng.lyu@utah.edu}
\author{Yue Zhao}
\email{zhaoyue@physics.utah.edu}
\affiliation{Department of Physics and Astronomy, University of Utah, Salt Lake City, UT 84112, USA}

\begin{abstract}

The QCD axion is a well-motivated hypothetical particle beyond the Standard Model (SM) and a compelling dark matter candidate. Its relic abundance is highly sensitive to the thermal history of the universe when the temperature is around the QCD confinement scale. Meanwhile, the NANOGrav Collaboration has reported evidence for a stochastic gravitational wave background, which could originate from a supercooled first-order phase transition (FOPT) with a nucleation temperature around the O(MeV-GeV) scale. We explore how such an FOPT might alter the evolution of the QCD axion. Our findings suggest that it could induce the axion to go through a short stage of mini kinetic misalignment. Moreover, in some parameter regime, the formation of QCD axion domain walls becomes generically expected. This has intriguing implications for both the existence of the QCD axion and the FOPT interpretation of the NANOGrav signal.

\end{abstract}

\maketitle

\setcounter{secnumdepth}{3}
\setcounter{tocdepth}{1}


{\bf Introduction.}---%

The QCD axion provides an elegant solution to the strong CP problem by promoting the QCD $\theta$ angle to a dynamical field~\cite{Peccei:1977hh,Peccei:1977ur,Weinberg:1977ma,Wilczek:1977pj,Shifman:1979if,Kim:1979if}. 
As the pseudo-Nambu-Goldstone boson of a spontaneously broken global $U(1)_{\text{PQ}}$ symmetry at the scale $f_a$, the QCD axion emerges as a compelling dark matter candidate through the misalignment mechanism.
Initially frozen at high temperature, the axion field begins coherent oscillations when the Hubble parameter $H(T)$ becomes comparable to its temperature-dependent mass $m_a(T)$. 
The axion mass exhibits a strong power-law dependence on the thermal bath temperature, $m_a(T) \propto T^{-4}$, making its relic abundance particularly sensitive to the thermal history of the universe near the QCD confinement scale, i.e. at $T_{\text{QCD}}\sim  \Lambda_{\rm QCD}$. 
Any deviation from standard cosmology during this epoch must therefore be carefully evaluated for its impact on the axion's evolution.

Recently an excess has been reported by various Pulsar Timing Array (PTA) datasets~\cite{NANOGrav:2023gor,EPTA:2023fyk,Reardon:2023gzh,Xu:2023wog}. 
The NANOGrav Collaboration has published the study~\cite{NANOGrav:2023pdq,NANOGrav:2023hvm} on the supermassive Black Hole binaries as the possible explanation of the data, as well as
alternative new physics cosmological scenarios.  
In the same study, it is shown that data  favors the cosmological origin, and the
FOPT~\cite{Bringmann:2023opz,Chen:2023bms,Wang:2023bbc,Ghosh:2023aum,Ratzinger:2020koh,Bringmann:2023opz,Croon:2024mde,Winkler:2024olr,An:2023jxf,Zheng:2024tib,Chen:2023bms,He:2023ado,Ghosh:2023aum} emerges as one of the most appealing choices. 

To account for the significant signal strength, the FOPT must be sufficiently strong and occur at a relatively low temperature. 
This motivates a strong supercooling scenario 
that has been extensively studied in the literature~\cite{Addazi:2023jvg,Balan:2025uke,Costa:2025csj,Goncalves:2025uwh,Goncalves:2024lrk,Gouttenoire:2023bqy}. 
The nucleation temperature can range from MeV to around QCD confinement scale, leading to a much higher reheating temperature $T_{\rm reh}$ after the FOPT.

The deviation from standard thermal history suggested by the PTA excess motivates a careful study of its implications for the QCD axion cosmology. In this work, we investigate the evolution of the QCD axion in the context of a supercooled first-order phase transition with a reheating temperature higher than $\Lambda_{\text{QCD}}$.

Interestingly, the strong temperature dependence of the axion mass can lead to a short period of time when the axion field develops non-trivial field-space velocity, enabling it to overcome potential barriers and undergo multiple cycles in its potential. This has important consequences for the axion relic abundance. Furthermore, the stochastic nature of bubble nucleation~\cite{Jinno:2016vai,Elor:2023xbz} results in spatial variations in reheating temperatures among different bubbles. Such inhomogeneities can generate different axion evolution histories, potentially giving rise to domain wall (DW) formation in the absence of cosmic strings. We will explore these phenomena in this study.

{\bf Axion Evolution in the non-standard cosmology.}---%

In our study, we focus on the case where the PQ symmetry breaking happens before the inflation. Thus all Hubble patches share the same initial axion field angle ($\theta_{a,i}$) to start with. 
In the following, we consider two scenarios. 

\noindent \emph{Scenario A: } Inflaton field dominantly decays to SM particles after inflation. The early universe consists of both the SM thermal bath and the potential energy in a dark scalar field $\phi$ which triggers the FOPT around MeV $\sim$ GeV.

\noindent\emph{Scenario B: }Inflaton field dominantly decays to the dark sector after inflation. The early universe consists of both dark radiation and the potential energy in a dark scalar field $\phi$.

In both scenarios, as the universe expanses, the dark scalar field $\phi$ will dominate the energy density for a short period of time and an FOPT occurs which reheats the SM thermal bath. 
The key distinction between the two scenarios lies in the earlier temperature of the SM sector. This difference results in distinct evolutions of the QCD axion field due to its temperature dependent mass. Initially, the axion is frozen thanks to the Hubble friction, and it begins to roll once its mass becomes comparable to the Hubble parameter. 
Let us define the axion oscillation temperature $T_{\rm osc}$ as
\begin{equation}
\left.
\begin{matrix}
    A: \quad &m_{a,0} \brr{\dfrac{T_{\rm osc}}{T_{\rm QCD}}}^{-4} \\
    B: \quad &m_{a,0} \\
    \end{matrix}
\right\}
    =  m_a(T_{\rm osc})  \simeq 3H_{\rm osc}.
\end{equation}
Here $m_{a,0}$ is the zero temperature axion mass, and $H_{\rm osc} = \sqrt{\rho_\Lambda+\rho_{\rm rad}}/M_{\rm pl}$, with $\rho_\Lambda$ as the energy density stored in the dark scalar field $\phi$ and $\rho_{\rm rad}$ as the radiation energy density. 

\begin{figure}
    \centering
    \includegraphics[width=0.5\linewidth]{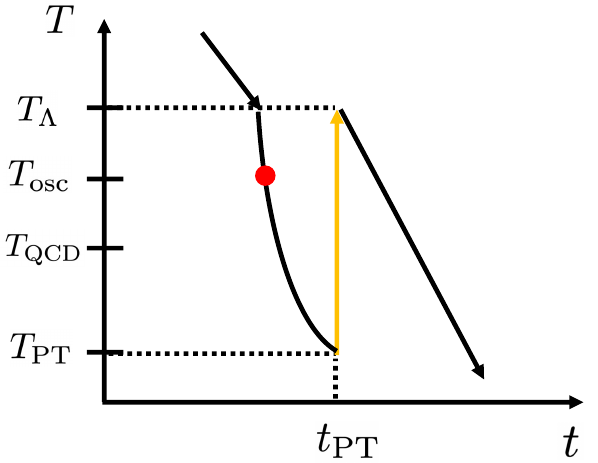}
    \caption{An illustration for the radiation temperature evolution. }
    \label{fig:illustration}
\end{figure}

At early times, the universe is SM radiation dominated in Scenario A and dark radiation dominated in Scenario B. 
In the aftermentioned maintext, we will use radiation to cover both case and not distinguish them. 
As the universe continues to expand, it reaches a $\Lambda$-radiation equality epoch at temperature $T_\Lambda$, determined by the condition:$\rho_\Lambda = \pi^2 g_*(T_\Lambda) T_\Lambda^4/30$.
Here $g_*(T_\Lambda)$ represents the effective relativistic degrees of freedom at temperature $T_\Lambda$ for either the SM or the dark sector.
This equality marks the transition when the potential energy of the dark sector begins to dominate over the radiation energy density as shown in Fig.~\ref{fig:illustration}. 

In Scenario A, where the axion mass grows rapidly with the expansion of the universe, coherent oscillations may start either before or after the temperature scale $T_\Lambda$. In contrast, for Scenario B, where the axion mass remains constant, we assume the axion oscillation occurs prior to $T_\Lambda$.

An illustration on the evolution of the universe is presented in Fig.~\ref{fig:illustration}. We note that $T_{\rm osc}$ is not necessarily lower than $T_\Lambda$, and the  temperature for the FOPT, i.e. $T_{\rm PT}$, does not have to be lower than $T_{\rm QCD}$. 
Once the axion starts to oscillate, its amplitude is reduced due to the universe expansion. For Scenario A, additional reduction is caused by the increased axion mass. In this case, using the WKB approximation,
the ratio of the axion number density to the entropy density is conserved~\cite{Fox:2004kb}. 
Including all these factors, we can write the axion field at the moment of FOPT as
\begin{equation}\label{eq:ini_theta}
\begin{split}
    &\theta_{a, {\rm PT}}=\theta_{a}(t = t_{\rm PT}) \simeq \theta_{a,i} \brr{\dfrac{m_a(T_{\rm osc})}{m_{a}(T_{\rm PT})}}^{1/2} \\
    &  \brr{\dfrac{T_{\rm PT}}{T_{\rm osc}}}^{3/2} \brr{\dfrac{g_{*S}(T_{\rm PT})}{g_{*S}(T_{\rm osc})}}^{1/2} \cos\phi_{\rm PT} \ .
\end{split}
\end{equation}
Here $\phi_{\rm PT}$ is the axion oscillating phase at $t = t_{\rm PT}$. For Scenario B, the axion mass ratio in Eq. \ref{eq:ini_theta} is simply 1 because the axion mass is always the zero-temperature value $m_{a,0}$ before the FOPT. In addition, the ratio of $g_{*S}$ is assumed to be 1 in the dark sector.
Furthermore, the axion  velocity in the field space can be written as
\begin{equation}\label{eq:velocity}
\begin{split}
 \dot \theta_{a, {\rm PT}}=& \dot{\theta}_{a}(t_{\rm PT} )  \simeq   - \theta_{a,i} \srr{m_{a}(T_{\rm PT}) \, m_{a}(T_{\rm osc})}^{1/2} \\
    & \brr{\dfrac{T_{\rm PT}}{T_{\rm osc}}}^{3/2} \brr{\dfrac{g_{*S}(T_{\rm PT})}{g_{*S}(T_{\rm osc})}}^{1/2}  \sin \phi_{\rm PT} \ .
\end{split}
\end{equation}

In both scenarios, the Standard Model sector is reheated following the FOPT. In Scenario A, the reheating temperature coincides with $T_\Lambda$. For Scenario B, we assume comparable relativistic degrees of freedom ($g_*$) in both sectors, such that $T_\Lambda$ remains a valid approximation for the reheating temperature. 
The raise of the temperature in the SM sector suddenly flattens the axion potential. This allows the kinetic energy to drive the axion field to undergo circular motion in the nearly flat potential. Following the convention of \cite{Co:2019jts,Co:2020dya,Eroncel:2024rpe}, we call such a phenomenon as \emph{mini kinetic misalignment}.

Right after reheating, the axion kinetic energy density is
$K(T_{\Lambda}) =  \dot{\theta}_{a,{\rm PT}}^2 f_a^2/2$,
and the hight of the potential energy barrier is given by $ V_{\rm max}(T_{\Lambda}) = 2 f_a^2 m_a(T_\Lambda)^2$.
Mini kinetic misalignment occurs if the kinetic energy is much larger than the potential barrier, 
\begin{equation}\label{eq:kin_condtion}
    K(T_{\Lambda}) \gg V_{\rm max}(T_{\Lambda}) \  .
\end{equation}
With the expansion of the universe, the axion kinetic energy evolves as $a^{-6}$ and the axion potential would be lifted up, evolving as $a^{8}$. 
This leads to the trapping temperature $T_{\rm trap}$, when the kinetic energy is comparable to the potential, namely, $K(T_{\rm trap}) \simeq V_{\rm max}(T_{\rm trap})$. This gives $T_{\rm trap}/T_\Lambda = \brr{K(T_\Lambda)/V(T_\Lambda)}^{1/14}$. 
Such high power dependence makes the axion quickly be trapped unless the ratio $K(T_\Lambda)/V_{\rm max}(T_\Lambda)$ is extremely high.

{\bf Axion Field Evolution - Homogenous Background}-
Firstly let us assume the universe gets reheated only after the FOPT completes. In this case, the reheating temperature is approximately universal in space, and the axion field evolves as a homogeneous background. 
The equation of motion (EOM) of the axion after reheating is given by 
\begin{equation}\label{eq:axion_eom}
\Ddot{\theta}_a + 3 H(t) \dot{\theta}_a + m_a^2(t) \sin\theta_a = 0 \  ,
\end{equation}
in which the Hubble scales as $H(t) \simeq (2t)^{-1}$ and the axion mass increases as the SM temperature decreases prior to the QCD confinement scale. 
Unlike the ordinary axion misalignment calculation, 
the axion field has a non-trivial initial velocity as $\dot{\theta}_{a,{\rm PT}}$. 
The evolution of the system can be classified into the kination and trapped regimes. 
If Eq.~(\ref{eq:kin_condtion}) is satisfied, 
the axion potential is negligible, and the solution can be approximately written as
\begin{equation}\label{eq:theta_evo}
    \theta_a(t) = \theta_a(t_{\rm PT}) + \dfrac{\dot{\theta}_{a,{\rm PT}}}{H_\Lambda} \srr{1 - \brr{\dfrac{t_{\rm PT}}{t}}^{1/2}}  \  .
\end{equation}
This equation applies until the axion is trapped namely $K(t)/V_{\rm max}(t)$ drops down to around 1. 
This indicates that the axion rotation is dominantly damped by the Hubble friction, and the $\theta_a$ value changes significantly over O(1) Hubble time. 
The alternative case is $K(T_{\rm PT})/V_{\rm max}(T_{\rm PT}) <1$. In this case, the kinetic energy is negligible and the mini kinetic misalignment is not effective. The comparison between the Hubble expansion $H_\Lambda$ and the axion mass $m_a(T_\Lambda)$ determines the axion's evolution. 
If $H_\Lambda \ll m_a(T_\Lambda)$, the axion field oscillates around the potential minimum and it dilutes as matter. 
Conversely, if $H_\Lambda \gg m_a(T_\Lambda)$, the axion is overdamped and quickly frozen. It resumes oscillation once the Hubble rate decreases sufficiently. 

\begin{figure}
    \centering
    \includegraphics[width=0.45\linewidth]{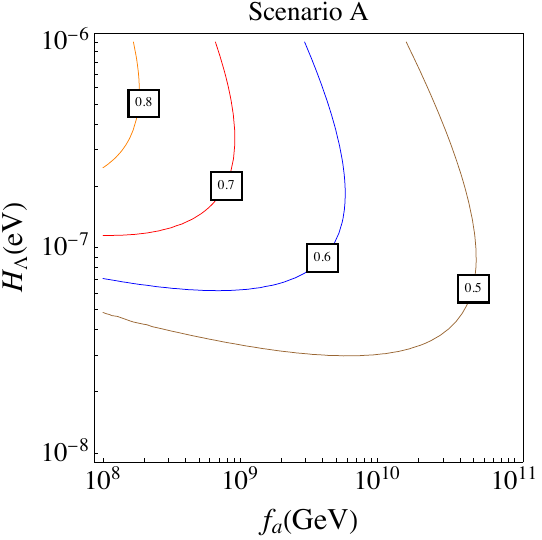}
    \includegraphics[width=0.45\linewidth]{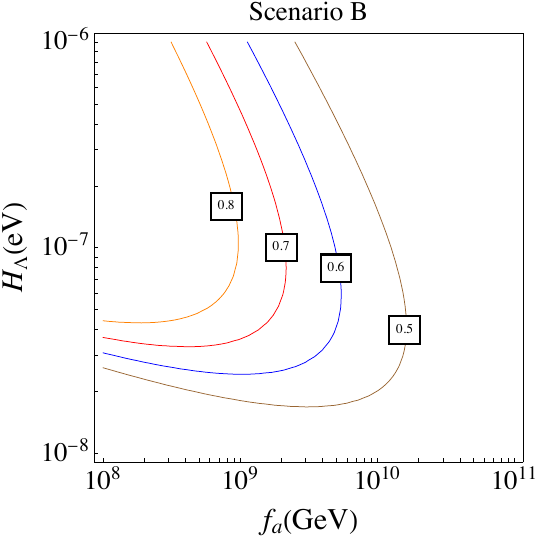}
    \caption{The contours of the reset $\theta_a$ value as Eq. (\ref{eq:theta_evo}) by taking $t$ to be the trapping moment. Here we choose $\theta_{a,i} = 0.2$ and $\sin\phi_{\rm PT} = 1/\sqrt{2}$.}
    \label{fig:reli_alter}
\end{figure}

The existence of the mini kinetic misalignment era can alter the DM relic abundance. As a demonstration, we choose the benchmark with the initial axion misalignment as $\theta_{a,i} = $  0.2 and $\sin\phi_{\rm PT}$ is averagely set to be $1/\sqrt{2}$. 
We solve Eq.~(\ref{eq:axion_eom}) explicitly and calculate the subsequent maximum $\theta_a$ value as the  ``\emph{reset}" initial misalignment value. 
In Fig.~\ref{fig:reli_alter}, we show the contours of the reset axion field value in Scenario A and B.
Four contours with the \emph{reset} $\theta_{a}$ values equaling to 0.5,0.6,0.7 and 0.8 are displayed. The parametric regime with lower $f_a$ and higher $H_\Lambda$ values can cause larger reset misalignment value. For larger $\theta_{a,i} \sim O(1)$ value, the axion field can rotate over more than one cycle before being trapped. 
%

{\bf Axion Field Evolution - Inhomogeneity from Bubbles}-
The stochastic nature of the bubble formation during the FOPT can lead to inhomogeneity to the axion evolution. 
The nucleation rate of bubbles can be expressed as $\Gamma(t) \sim \Gamma_{\rm PT} \exp(\beta(t-t_{\rm PT}))$, where $\beta = d S_4/dt$ characterizing the duration of the phase transition. 
Since bubble nucleation is inherently stochastic, bubbles can form at slightly different times across different spatial regions. As a result, there is a relatively broad distribution in the spatial and temporal profiles of bubbles before they begin to merge. 
In~\cite{Turner:1992tz}, the authors have calculated this distribution which is peaked at $\beta^{-1}$ with the bandwidth $O(0.1) \beta^{-1}$. 
If the interaction between the dark sector and the SM sector is sufficiently weak, the universe may require a prolonged period to reheat the SM sector after the  FOPT, taking longer than the time needed for all bubbles to merge. 
In this case, the axion field evolves identically on both sides of the bubble walls, resulting in a homogeneous axion distribution, as discussed previously. 
Conversely, if the dark sector reheats the SM sector before the bubbles fully merge, an inhomogeneous axion background will be generated.

More specifically, within the bubbles, the scalar potential energy is converted into the SM thermal bath. In particular, if the reheating temperature exceeds the QCD confinement scale, the axion potential becomes significantly flattened inside the bubble. This causes a distinct evolution of the axion field on two sides of the bubble wall.
The dynamics of the axion field in the bubble is determined by solving its equation of motion, 
\begin{equation}\label{eq:eom}
    \Ddot{\theta}_a + 3 H \dot{\theta}_a - \dfrac{\nabla^2}{a(t)^2} \theta_a + m_a^2(T)\sin\theta_a = 0.
\end{equation}
Particularly, the continuity requirement sets the boundary condition at the bubble wall. 

In the following, we consider two representative reheating scenarios:
(i) a feeble coupling case, where the reheating timescale is comparable to the duration of the FOPT, and
(ii) the instantaneous reheating limit, where the SM sector is reheated almost immediately after bubble nucleation. 
As discussed above, the Scenario A and B only lead to a difference in terms of the initial condition when solving Eq. (\ref{eq:eom}). The qualitative behavior is very similar.

\underline{Feeble Coupling}:
We consider the scenario in which the SM thermal bath is reheated to $T_\Lambda$ when a bubble is about to merge with others, i.e. when the bubble size $r_s$ becomes $O(\beta^{-1})$. Under the assumption of bubble wall velocity as the speed of light, 
the axion mass can be written as 
\begin{equation}\label{eq:bubble_boundary_cond}
 m_a(r_s + \delta t,r) =    
\begin{cases}
 m_a(T_\Lambda) & r<r_s + \delta t\\
 m_{a}(T_{\rm PT}) 
 & r\geq r_s + \delta t
\end{cases}
\end{equation}
This sets the initial condition when solving for Eq. (\ref{eq:eom}).

Deep in the bubble interior, the axion rotates in the field space, driven by the initial velocity. 
Due to continuity requirement, the axion field value at the bubble wall continues to oscillate around $ \theta_a=0$. 
As the bubble expands, such a boundary condition induces a spatial gradient inside the bubble.
Consequently, the oscillatory pattern at the boundary propagates inward, thereby influencing the evolution of the axion field throughout the bubble interior.
\begin{figure}[t]
    \centering
    \includegraphics[width=0.7\linewidth]{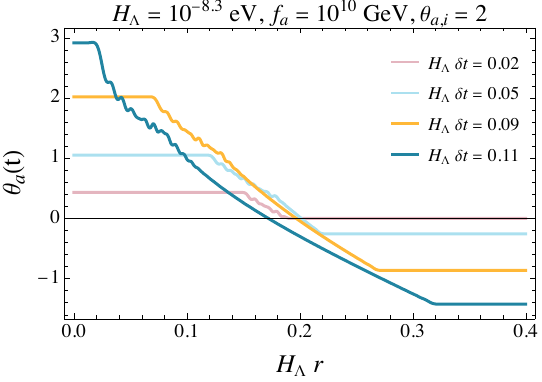}
    \caption{The axion profiles at different time in the feeble coupling case.}
    \label{fig:feeble-profile}
\end{figure}
As an illustration, we numerically solve Eq. (\ref{eq:eom}) and present the spatial profile of the axion field at several representative time slices in Fig.~\ref{fig:feeble-profile}. Here we choose a benchmark value $r_s = 0.17 H_\Lambda^{-1}$ and $\beta/H_\Lambda = 5$. 
The value of $r_s$ is chosen to be smaller than $O(\beta^{-1})$ so that when bubbles merge, the internal region has been reheated.
At $t = t_s$, the axion velocity is identical in both the interior and exterior regions of the bubble. However, due to the differences in the potentials, the field kinematics gradually deviate between the two regions.
This leads to the development of spatial gradients across the intermediate transition region.
The induced fluctuations propagate at the speed of light.
The axion in the inner flat region keeps rolling with the velocity $\dot{\theta}_a(t_s)$. 
The boundary effect reaches the bubble center at $\delta t = r_s$.

\begin{figure}[t]
    \centering
    \includegraphics[width=0.75\linewidth]{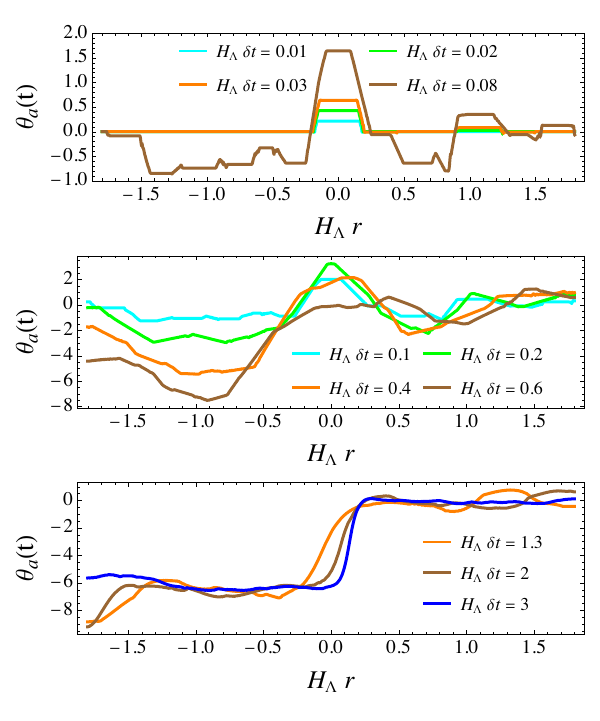}
    \caption{The axion profiles evolution in the feeble coupling case. Here we simulate multiple bubbles nucleation and merging. The DW forms as the axion potential barrier height is lifted up in the lower panel. }
    \label{fig:axion-feeble-evo}
\end{figure}

After bubbles fully merge with themselves,  the entire universe transitions into the true vacuum with reheated temperature.
Modeling the full three-dimensional dynamics of bubble merging is highly nontrivial and lies beyond the scope of this work. Instead, we perform a 1D simulation which  qualitatively captures all important physics.
In Fig.~\ref{fig:axion-feeble-evo}, we show the nucleation and merging of multiple bubbles. 13 bubbles are randomly nucleated within the spatial range $[-1.8,1.8] H_\Lambda^{-1}$. 
We choose the same parameters as in Fig.~\ref{fig:feeble-profile}.
It is assumed that the bubble centered at the origin is nucleated first and other bubbles are nucleated later on within a short period of time. Here 
$\delta t = 0$ corresponds to the moment when the interior region of the first nucleated bubble is reheated. We assume that, after the moment of $\delta t = 0.08 H_\Lambda^{-1}$, the FOPT is complete and the whole universe is converted to the high temperature phase. 
The upper panel of Fig.~\ref{fig:axion-feeble-evo} shows the evolution of the axion field before the FOPT completes, i.e. from $\delta t = 0.01 H_\Lambda^{-1}$ to $\delta t = 0.08 H_\Lambda^{-1}$. 
The axion field inside the bubble rolls with a velocity determined by  Eq.~\ref{eq:velocity}. 
The regions outside the bubbles remain in the low-temperature phase, keeping the axion field close to zero. Consequently, the axion field configuration evolves as a combination of flat regions and slopes. These sharp features persist in the early stages because the axion in the high temperature regions is nearly massless, ensuring that all modes obey the dispersion relation $\omega = k$. As a result, any linear superposition of these solutions remains a valid solution to the equation of motion. 

As time progresses, the bubbles begin to merge, and the middle panel of Fig. \ref{fig:axion-feeble-evo} shows the subsequent evolution of the axion field from $\delta t = 0.1 H_\Lambda^{-1}$ to $0.6 H_\Lambda^{-1}$. Compared to the upper panel, the sharp features persist for some time but gradually smooth out at later times. This occurs because, as the temperature decreases, the axion potential term grows increasingly significant, causing the linear dispersion relation to become a poor approximation.
Finally, the lower panel illustrates the evolution of the axion field once the potential barrier exceeds the kinetic energy of the field. At this stage, the axion field in different regions becomes trapped in distinct domains, resulting in the formation of axion DWs.

One can qualitatively estimate the conditions for domain wall formation as follows. For each bubble, the amplitude of the axion field at its center is given by the product of two factors: (1) the axion field velocity at reheating $\dot{\theta}_a = m_{a}(T_{\rm PT}) \, \langle\theta_{a}\rangle$ where $\langle\theta_{a}\rangle$ represents the typical oscillation amplitude after averaging on phases, and (2) the timescale before bubble merging, which is approximately $\beta^{-1}$. 
Additionally, accounting for Hubble expansion, the amplitude of a massless field redshifts linearly with temperature. 
From bubble formation until the axion becomes trapped, this redshift factor is approximately $(T_{\rm trap}/T_\Lambda)$. Combining these factors, we obtain the qualitative condition for axion domain wall formation as 
\begin{equation}\label{eq:DW_cond}
    \dfrac{m_{a}(T_{\rm PT}) \, \langle\theta_{a}\rangle}{\beta} \brr{\dfrac{T_{\rm trap}}{T_\Lambda}}
    > \pi \  .
\end{equation}
%

\underline{Instantaneous Reheating}---%
We now turn to an alternative scenario in which the decay of dark sector particles into the Standard Model sector is sufficiently rapid that reheating can be treated as instantaneous.
In this case, the initial condition of the axion field is effectively set at the moment of bubble nucleation, corresponding to setting $r_s \to 0$ in Eq. (\ref{eq:bubble_boundary_cond}).
Under this boundary condition, and neglecting both Hubble friction and the axion mass $m_a(T_\Lambda)$, the axion field profile can be obtained analytically in three-dimensional spherical coordinates. 
\begin{figure}
    \centering
    \includegraphics[width=0.75\linewidth]{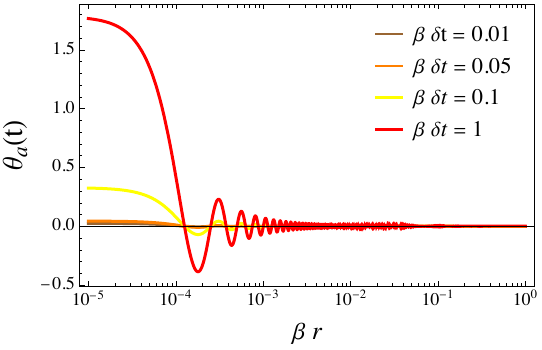}
    \caption{The axion field angle profile at different moments. Here we assume $m_{a,0}/\beta = 5 \times 10^{4}$, $\langle\theta_a\rangle = 10^{-4}$ 
    .}
    \label{fig:InstantProfile}
\end{figure}
Fig.~\ref{fig:InstantProfile} shows the axion field profile at different moments. 
One can verify that the oscillation amplitude decreases with radius as $r^{-1}$.
At the bubble center $r = 0$, the oscillation amplitude grows linearly with time $\delta t$.
Similar to the previous case, the peak amplitude at the bubble center reaches a value comparable to $ m_{a}(T_{\rm PT}) \langle\theta_{a}\rangle/\beta$ before bubble merging. 
There are two distinct features compared to the feeble coupling case. First, the dominant fluctuation corresponds to the mode with $k \sim m_{a,0}/2$, which is triggered by the axion oscillation at the bubble wall boundary. 
In addition, the oscillation amplitude inside the bubble spatially decreases as $r^{-1}$. 
As a result, the significant deviation from 0 only concentrates near the bubble center, within $O(m_{a,0}^{-1})$ range. 
The amplitude at the center redshifts as $a^{-1}$.
The condition for the DW formation in this instantaneous reheating case is again given by Eq.~(\ref{eq:DW_cond}). 
In contrast to the feeble coupling case, due to the narrow width of the fluctuation, the size of the region enclosed by DWs tends to be very small. 

\begin{figure}
    \centering
    \includegraphics[width=0.47\linewidth]{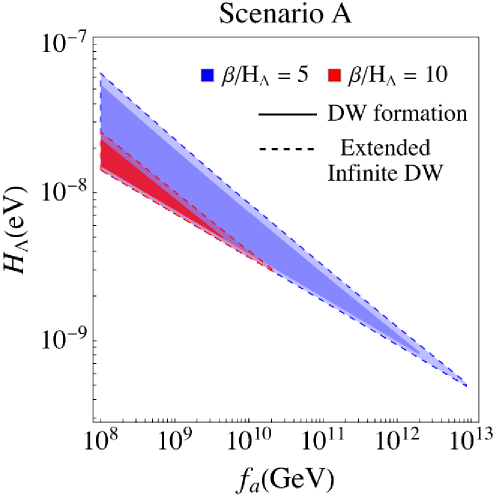}
    \includegraphics[width=0.47\linewidth]{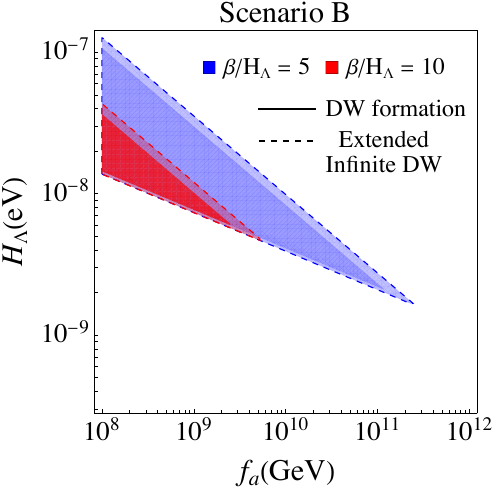}
    \caption{The parametric space that can produce the DWs on $H_\Lambda-f_a$ plane according to Eq. (\ref{eq:DW_cond}). The left panel is for Scenario A and the right panel is for Scenario B. Here we choose $\theta_{a,i} = 2$.  The light region with dashed boundary refers to infinite extended DWs formation in feeble coupling case and the solid region represents the DW formation in both feeble coupling and instantaneously reheating cases.}
    \label{fig:DW-para}
\end{figure}
In Fig.~\ref{fig:DW-para}, we display the relevant parameter space on  $H_\Lambda-f_a$ plane to produce the DWs in both feeble coupling case and instantaneous reheating case. 
The two colored regions are for $\beta/H_\Lambda = 5$ and 10.
Here we assume $\theta_{a,i} = 2, T_{\rm PT} = T_{\rm QCD}$ for Scenario A and $T_{\rm PT} = T_\Lambda/2$ for Scenario B. The condition $K(T_\Lambda)/V_{\rm max}(T_\Lambda) > O(1)$ 
is imposed to ensure that the axion has sufficient kinetic energy to overcome the potential barrier. 
The lighter shaded region is the parameter space where the DW can form. The darker shade represents the region where an extended infinite DW may form when the coupling is feeble. We will comment on this subtlety in the next section.

{\bf Axion DW Formation and Evolution}---%

We note that the mini kinetic misalignment discussed in this study is a relatively low energy process. There are no cosmic strings produced, and consequently the DWs being induced are not bounded.  
Such DWs can be classified into two types, the finite enclosed DWs and the infinitely extended DWs~\cite{Vilenkin:2000jqa}. 
The finite enclosed DWs are expected to form locally with a size smaller than the Hubble patch. 
They are expected to shrink under the surface tension and eventually convert to axion radiation via collapse and vibration. 

The infinitely extended DWs lead to cosmological disasters. Such extended DWs evolve rapidly into the scaling regime~\cite{Garagounis:2002kt,Leite:2011sc}, with approximately one extended DW per Hubble patch. The stable DW energy density is given by $\rho_w \sim \sigma/t$ with $\sigma$ the surface tension. It scales as $a^{-2}$ in the radiation-dominated era and redshifts slower than the radiation or matter components, thus eventually dominate the universe. The current observation sets the DW network surface tension $\sigma$ below 1 $\MeV^3$~\cite{Zeldovich:1974uw}, which is far smaller than the tension of the QCD axion, i.e. $m_{a,0} f_a^2 \sim \Lambda_{\rm QCD}^2 f_a$. This means that the QCD axion DWs once formed will eventually dominate our universe, in contradiction to the current observation.

The probability of extended infinite DWs formation can be given by the percolation theorem. For the scenario we are considering, since axion in each bubble randomly chooses the positive or negative direction to evolve when the reheating happens, it is very similar to the $\mathbb{Z}_2$ spontaneous symmetry breaking in a three dimensional space. The critical value for the volume ratio of choosing the two vacua in order to generate an extended infinite DW is $p_c = 0.31$~\cite{STAUFFER19791}. 
Since the axion mass before reheating is much larger than the Hubble scale during the FOPT, the axion oscillation phases among bubbles can be well approximated as uniformly distributed.  Under this approximation, one can determine the parameter space in which extended infinite DWs form, corresponding to the darker shaded region in Fig.~\ref{fig:DW-para}. 
On the other hand, as we have argued above, the DWs formed in the instantaneous reheating case is very small in size, approximately $m_{a,0}^{-1}$. Therefore the extended infinite DWs cannot form in that limit.

{\bf Conclusion and Outlook}---%
The NANOGrav signal strongly motivates the consideration of a FOPT around QCD scale in the dark sector, which reheats the Standard Model thermal bath and  impact the evolution of the QCD axion. Given the fact that the QCD axion potential is highly sensitive to the SM thermal temperature, such a FOPT may induce a mini kinetic misalignment, causing the axion field to undergo rotation in field space. This scenario has two immediate consequences. First, it resets the value of axion field misalignment, disrupting the conventional relationship between the initial misalignment angle and relic abundance. Second, it leads the formation of domain walls even in the absence of cosmic strings. These domain walls can be finite-sized enclosed walls which collapse into axion radiation and further correct the relic abundance, or infinitely extended domain walls entering a scaling regime which poses severe cosmological constraints. In this paper, we perform a first analysis to study the parameter space where the mini kinetic misalignment may happen and what phenomena it may lead to.

Beyond these effects, the interplay between FOPT at QCD scale and the QCD axion physics opens new directions to explore. These include gravitational wave signatures from domain wall evolution and potential primordial black hole~\cite{Vachaspati:2017hjw,Ferrer:2018uiu,Ge:2019ihf,Ge:2023rrq,Dunsky:2024zdo} formation from collapsing domain walls. The future study on these aspects could further refine our understanding of axion cosmology in the context of phase transitions.

{\bf Acknowledgment}\label{sec:acknowledgement}
We thank Raymond Co and Shuailiang Ge for useful discussions. K-F.L and Y.Z. are supported by U.S. Department of Energy under Award No. DESC0009959.

\bibliography{ref}

\end{document}